\documentclass[11pt,a4paper]{article}

\usepackage{jheppub}
\usepackage{physics}
\usepackage{empheq}
\usepackage{color}
\usepackage{tensor}
\usepackage[whole]{bxcjkjatype}
\usepackage{url}
\usepackage{cleveref}
\usepackage{booktabs}

\newcommand{\del}{\partial}
\newcommand{\Slash}[1]{{\ooalign{\hfil/\hfil\crcr$#1$}}} 
\newcommand{\nn}{\nonumber\\}
\newcommand{\p}{\partial}

\newcommand{\df}{\text{d}}
\newcommand{\sfdv}[3]{\frac{\delta^2 #1}{\delta #2\delta #3}}

\newbox{\ORCIDicon}
\sbox{\ORCIDicon}{\large \includegraphics[width=0.8em]{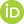}}

\begin{document}

\begin{flushright}
KUNS-3005
\end{flushright}

\title{Functional Renormalization Group Analysis of $O(3)$ Nonlinear Sigma Model and Non-Abelian Bosonization Duality}

\author[a]{Junichi Haruna\,\href{https://orcid.org/0000-0002-1828-8183}{\usebox{\ORCIDicon}}}

\author[b]{, Keito Shimizu\,\href{https://orcid.org/0009-0003-5200-6092}{\usebox{\ORCIDicon}}}

\author[c]{, and Masatoshi Yamada\,\href{https://orcid.org/0000-0002-1013-8631}{\usebox{\ORCIDicon}}}

\affiliation[a]{Center for Quantum Information and Quantum Biology, Osaka University, Toyonaka, Osaka 560-0043, Japan}

\affiliation[b]{Department of Physics, Kyoto University, Kitashirakawa-Oiwakecho, Kyoto 606-8502, Japan}

\affiliation[c]{Center for Theoretical Physics and College of Physics, Jilin University, Changchun 130012, China}

\emailAdd{j.haruna1111@gmail.com}
\emailAdd{kate@gauge.scphys.kyoto-u.ac.jp}
\emailAdd{yamada@jlu.edu.cn}

\abstract{
It is known that the $U(2)$ Wess-Zumino-Witten model is dual to the free fermion theory in two dimensions via non-Abelian bosonization. 
While it is decomposed into the $SU(2)$ Wess-Zumino-Witten model and a free compact boson, the former is believed to be equivalent to the $O(3)$ nonlinear sigma model with the theta term at $\theta=\pi$.
In this work, we reexamine this duality through the lens of non-perturbative renormalization group (RG) flow.
We analyze the RG flow structure of the $O(3)$ nonlinear sigma model with the theta term in two dimensions using the functional renormalization group.
Our results reveal a nontrivial fixed point with a nonzero value of the topological coupling.
The scaling dimensions (critical exponents) at this fixed point suggest the realization of a duality between the $O(3)$ nonlinear sigma model with the theta term and the free fermion theory, indicating that these models belong to the same universality class.
}

\maketitle

\section{Introduction}
Quantum Field Theory (QFT) is the fundamental mathematical framework for describing the dynamics of quantum particles.
Since there is a limited number of exactly solvable systems, perturbation theory has been employed to capture certain nontrivial features in general models.
In particle physics, it has succeeded in elucidating the physics of the standard model.
However, as couplings veer from the weak-coupling regime, it fails to grasp the fundamental nature of the systems.
It is anticipated that strongly correlated systems such as quantum chromodynamics and the Hubbard model exhibit nontrivial aspects beyond the reach of perturbation theory.

When tackling nonperturbative phenomena in strongly interacting theories, we cannot rely on weak-coupling expansion and need another analytical method.
An approach that provides us an avenue for investigating the nonperturbative aspects of QFT is the use of duality. 
When we can map a strongly interacting system to another weakly interacting system, such that their correlation functions are equivalent, we refer to these systems as exhibiting strong-weak coupling duality.
If we have strongly/weakly interacting dual systems, nontrivial features arising from the dynamics in the former can be gleaned from the latter. 
A typical example of duality is the Anti-de Sitter (AdS)/Conformal Field Theory 
(CFT) correspondence~\cite{Maldacena:1997re,Witten:1998qj}.

Another example of duality is bosonization. 
In this context, composite states of fermionic degrees of freedom correspond to bosonic ones.
A well-known example is the duality between the massive Thirring model and the sine-Gordon model in two dimensions, as pointed out by Coleman~\cite{Coleman:1974bu}.
This correspondence is known as a kind of ``Abelian bosonization".
Furthermore, this idea was extended to the non-Abelian case, called ``Non-Abelian bosonization", by Witten~\cite{Witten:1983ar}.
It was shown in terms of the current algebra and CFT that massless free fermions with $N$-flavors are dual to the $U(N)$ Wess-Zumino-Witten model in two dimensions.
See also Ref.~\cite{Senechal:1999us} for review.

The $U(2)$ WZW model is particularly interesting, which can be decomposed into the $SU(2)$ WZW model and a free boson sector. 
This $SU(2)$ WZW model is believed to be dual to the $O(3)$ nonlinear sigma model with the theta term (or the Wess-Zumino term) with $\theta=\pi$.

This implies that the free fermion model with two-flavor, the $SU(2)$ WZW model, and the $O(3)$ non-linear sigma model with the theta term are closely related with each other.
Furthermore, the $O(3)$ nonlinear sigma model with the theta term is discussed as describing the infrared (IR) physics of the antiferromagnetic Heisenberg spin chain, which is related to the Haldane conjecture~\cite{Haldane:1982rj,Haldane:1983ru}.
Thus, investigating non-Abelian bosonization in the $N=2$ case is important not only for particle physics but also for condensed matter physics.

Our aim is to study the duality between the $O(3)$ nonlinear sigma model with the theta term and the free fermion model from the perspective of the nonperturbative renormalization group (RG).
To achieve this, we employ the Functional Renormalization Group (FRG)~\cite{Wilson:1971bg,Wilson:1973jj,Wegner:1972ih,Polchinski:1983gv,Wetterich:1992yh,Morris:1993qb,Reuter:1993kw,Ellwanger:1993mw,Morris:1998da,Berges:2000ew,Aoki:2000wm,Bagnuls:2000ae,Polonyi:2001se,Pawlowski:2005xe,Gies:2006wv,Delamotte:2007pf,Sonoda:2007av,Igarashi:2009tj,Rosten:2010vm,Braun:2011pp} as a powerful tool.
The FRG describes the change in the effective action under varying energy scales, expressed by a functional partial differential equation known as the flow equation.

The FRG method is thought to be suitable for studying duality for several reasons.
First, the derivation of the flow equations does not rely on the perturbative expansion of the couplings.
Second, flow equations can identify fixed points corresponding to conformal field theories (CFTs).
In particular, continuum quantum field theories are defined at fixed points of the RG flow, and possible interactions in the continuum theory are determined from the RG flow structure around these fixed points.
There are several references that analyze dualities with the FRG method~\cite{Nandori:2010ij, Daviet:2021whj,Zinn-Justin:1991ksq,Karkkainen:1993ef,Wilson:1971dc}.
While an FRG analysis of the $O(3)$ nonlinear sigma model was performed in Refs.~\cite{Flore:2012wh,Efremov:2021fub}, we reexamine it with a different FRG approach for the following reasons.

Applying the FRG to this model presents several subtle challenges.
It includes the theta term, which is topological, in addition to the kinetic term of the scalar fields.
Generally, topological terms do not contribute to the dynamics of a system within the FRG because their functional derivative vanishes.
As a result, these effects do not appear in the flow equations.
Another challenge arises from the nonlinearity of the $O(3)$ nonlinear sigma model, where the scalar fields $\phi_i$ are constrained by $(\phi_i)^2=1$, indicating that the target space of the field is $S^2$.
Although, naively, it is natural to respect this constraint $(\phi_i)^2=1$ along the RG flow, standard approximations in an FRG setup may violate it.

To address these issues, we follow the method used in Ref.~\cite{Fukushima:2022zor}.
We extend the target space of the scalar fields from $S^2$ to $\mathbb{R}^3$ in the formalism of the 1PI effective action, but not of the Wilsonian effective action.
This setup aims to extract a certain nontrival dynamics of the $O(3)$ {\it nonlinear} sigma model from the {\it linear} one.
Since the former model is the low-energy effective model of the latter model, these models should share a certain RG trajectory.
Therefore, we will start by studying the latter model as a first step to reveal the mechanism of the duality nonperturbatively.

This paper is organized as follows.
In \Cref{sec: review of bosonization}, we briefly review non-Abelian bosonization in two dimensions.
In particular, we focus on the $SU(2)$ case, so that we can derive duality between $SU(2)$ WZW model and $O(3)$ nonlinear sigma model with the theta term.
In \Cref{sec: FRG for O(3) linear WZW model}, we discuss several problems to handle the $O(3)$ nonlinear sigma model with the theta term within the FRG.
Then, we explain what kind of FRG setup is more suitable to address them.
In \Cref{sec: RG flow and fixed points}, we analyze the RG flow of the $O(3)$ nonlinear sigma model with the theta term, using the FRG.
We discuss the existence of a nontrivial fixed point under some approximations, in addition to the Gaussian fixed point.
We also study the scaling dimensions around these fixed points.
\Cref{sec: conclusion and discussion} is devoted to conclusions and discussions.

\section{Brief Review of Non-Abelian Bosonization in two dimension}
\label{sec: review of bosonization}
In this section, we briefly review the non-Abelian bosonization for free fermions in two-dimensional Euclidean spacetime~\cite{Witten:1983ar}.
Bosonization is a generic term for dualities that enables us to treat a theory of fermions as a theory of bosons.
The simplest example is the duality between the massless Dirac fermion theory and the free single massless scalar theory~\cite{Coleman:1974bu}:
\begin{align}
\mathcal L_\text{f} = \bar\psi(x) i\Slash\p \psi(x) \longleftrightarrow \mathcal L_\text{s} = \frac{1}{2}\p_\mu \phi(x) \p_\mu \phi(x),
\label{eq: Abelian bosonization}
\end{align}
where $\psi(x)$ is a two-component Dirac field, $\phi(x)$ is a real scalar field, $\Slash{\del}\coloneqq \gamma_\mu \del_\mu$ and $\gamma_\mu\,(\mu=1,2)$ are the Dirac matrices in two dimensions.
Here, the bosonic field is related to the fermionic fields such that
\begin{align}
J_\mu=\bar\psi(x) \gamma_\mu \psi(x) = \frac{1}{\sqrt{\pi}}\varepsilon_{\mu\nu}\p_\nu \phi(x),
\end{align}
and
\begin{align}
&\bar\psi(x) \psi(x) =  \mu \cos (\sqrt{4\pi}\phi(x)),&
&\bar\psi(x) i\gamma_5 \psi(x) = \mu \sin (\sqrt{4\pi}\phi(x)),
\end{align}
where $\gamma_5=\gamma_0\gamma_1$ is the chiral Dirac matrix, $\varepsilon_{\mu\nu}$ is the antisymmetric tensor and $\mu$ is some renormalization scheme dependent mass scale.
Note that in the two-dimensional spacetime, the mass dimensions of the fermionic and bosonic fields are given by $[\psi]=1/2$ and $[\phi]=0$, respectively.
Here, $J_\mu$ is the $U(1)$ current of the fermion theory.
In this sense, the transformation from $\mathcal L_\text{f}$ to $\mathcal L_\text{s}$ is called ``Abelian bosonization''.

This fact was extended by Witten~\cite{Witten:1983ar} to the case of the massless free Dirac fermion theory with $N$-flavor in two dimensions by means of the current algebra.
More specifically, the fermionic theory is described by the free action given by
\begin{align}
    S_{\mathrm{f}} = \int_{\mathbb{R}^2} \df^2x\, \Bar{\psi}_i(x)\Slash{\del}\psi_i(x),
    \label{eq: N flavor fermion theory}
\end{align}
where each $\psi_i(x)$ (for $i=1,\ldots,N$)  is a two-component Dirac field.
This fermionic action is bosonized to the level-1 $U(N)$ WZW model, whose action is given by
\begin{align}
    S_{\mathrm{WZW}} = \frac{1}{8\pi}\int_{\mathbb{R}^2} \df^2x \,\Tr(\del_{\mu}g\del_{\mu}g^{-1} )
    - \frac{i}{12\pi}\int_B \Tr(\Tilde{g}^{-1}\df\Tilde{g}\wedge\Tilde{g}^{-1}\df\Tilde{g}\wedge\Tilde{g}^{-1}\df\Tilde{g}),
    \label{eq: WZW model action}
\end{align}
where $g(x)$ is a $U(N)$-valued field, and $B$ denotes a three dimensional region whose boundary is given by $\mathbb{R}^2$, while $\Tilde{g}$ is also a $U(N)$-valued field on $B$ and is connected smoothly to $g(x)$ on $\mathbb{R}^2$ (i.e., $\eval{\Tilde{g}}_{\del B = \mathbb{R}^2} = g$). 
Here, $\df g$ is an exterior derivative of $g$ and $\wedge$ denotes the exterior product.
The second term in Eq.~\eqref{eq: WZW model action} is called the Wess-Zumino term. 
This duality states that expectation values for operators in fermionic theory~\eqref{eq: N flavor fermion theory} agrees with those for the corresponding operators in the bosonic theory~\eqref{eq: WZW model action}, that is, we have
\begin{align}
    \expval{O_{\mathrm{f}}(x_1)\cdots O_{\mathrm{f}}(x_n)}_{S_{\mathrm{f}}} = \expval{O_{\mathrm{WZW}}(x_1)\cdots O_{\mathrm{WZW}}(x_n)}_{S_{\mathrm{WZW}}}.
\end{align}
For instance, the correspondences between fermionic and bosonic operators are given as follows:
\begin{subequations}
\begin{align}
  (J_-)_{ij} = i(\psi_-)_i(\psi_-)_j &\sim  \frac{1}{2\pi}((\del_-g) g^{-1})_{ij}\,, \\
  (J_+)_{ij} = i(\psi_+)_i(\psi_+)_j &\sim  \frac{1}{2\pi}(g^{-1}\del_+g)_{ij}\,, \\
  (\psi_+)_i (\psi_-)_j &\sim \mu  g_{ij} \,, \\
  \Bar{\psi}\psi &\sim \mu \tr (g+g^\dagger)\,,
  \label{eq: duality of mass term}
\end{align}
 \label{eq:dictionary of non-Abelian bosonization}
\end{subequations}
where we have introduced the notation
\begin{align}
    &\psi_{\pm} = \frac{1\mp\gamma_5}{2}\psi\,,&
    &x_{\pm} = \frac{x_0\pm x_1}{\sqrt{2}}\,,&
    &\del_{\pm} = \frac{\del}{\del x_{\pm}}\,.
\end{align}
We call the relations \eqref{eq:dictionary of non-Abelian bosonization} by the ``dictionary of non-Abelian bosonization."
Note that the $U(N)$ WZW model can be divided into the $SU(N)$ model and the single massless scalar theory as follows. 
An $U(N)$-valued field $g(x)$ can be written by using an $SU(N)$-valued field $g'(x)$ and a compact boson field $\varphi(x)$ as
\begin{align}
\label{eq: decompose U(N) to U(1) and SU(N)}
    g(x) = e^{i\varphi(x)}g'(x).
\end{align}
Substituting this into the WZW action~\eqref{eq: WZW model action}, we have
\begin{align}
    S_{\mathrm{WZW}}^{U(N)} = S_{\mathrm{WZW}}^{SU(N)} + \frac{N}{2}\int \df^2x (\partial_{\mu}\varphi)^2.
\end{align}
From this expression, we can see that the single boson completely decouples from the $SU(N)$ WZW part.
Therefore, we ignore it in the following RG analysis unless we mention it.

\begin{figure}
    \centering
    \includegraphics[width=1.0\linewidth]{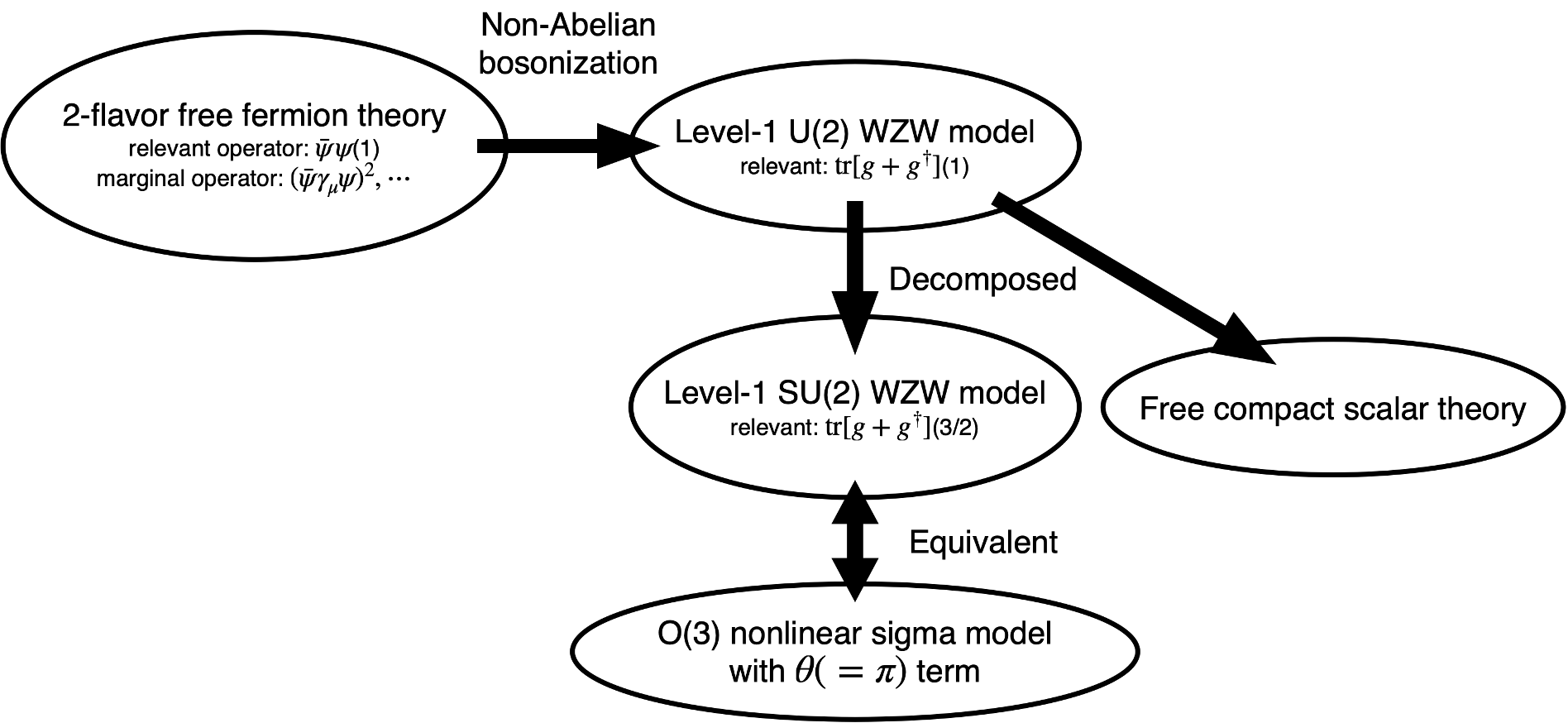}
    \caption{Models discussed in this paper and their relation. Numbers following relevant operators are their corresponding critical exponents. The critical exponents can be obtained by subtracting the operator dimensions from the spacetime dimension.}
    \label{fig:dualities}
\end{figure}

In this paper, we focus on duality in the case where $N=2$. It is known from the analysis of the antiferromagnetic spin chain that in the low energy region $SU(2)$ WZW model at the level $1$ is equivalent to the $O(3)$ nonlinear sigma model with the theta term~\cite{Affleck:1987ch}.
\begin{align}
\label{eq: NLt}
    S_{\mathrm{NL+top}} = \frac{1}{16\pi}\int_x\ \del_{\mu}\phi_i\del_{\mu}\phi_i + \frac{i\theta}{4\pi}\int_x\ \varepsilon_{\mu\nu}\varepsilon_{ijk}\phi_i\del_{\mu}\phi_j\del_{\nu}\phi_k \,,
\end{align}
with the constraint $\phi_i\phi_i=1$. Here and hereafter, we introduce the shorthand notation $\int_x = \int \df^2x$.
All models and their relation treated in this paper are depicted in \Cref{fig:dualities}.

Let us briefly see this equivalence\footnote{
    The following calculation is not a proof and just a naive discussion for explanation.
    If you want a more detailed discussion, see Section 8 of \cite{Altland:2006si} for example.
}.
An element $g$ of the Lie group $SU(2)$ can be parametrized by the scalar fields $\phi_i$ as 
\begin{align}
\label{eq: SU(2) field in terms of O(3) vector}
    g= \exp (i\frac{\sigma_i}{2} \phi_i(x))\,,
\end{align}
where $\sigma_i\ (i=1,2,3)$ are the Pauli matrices. If we consider only low-energy excitation states, we can expand $g$ in terms of $\phi_i$. Then, the kinetic term in the WZW model \eqref{eq: WZW model action} becomes
\begin{align}
    \frac{1}{8\pi}\int_{\mathbb{R}^2} \df^2x \Tr(\del_{\mu}g\del_{\mu}g^{-1})
    &= \frac{1}{16\pi}\int_{\mathbb{R}^2}\del_{\mu}\phi_i \del_{\mu}\phi_i + \cdots,
\end{align}
while the Wess-Zumino term in \Cref{eq: WZW model action} can be calculated as
\begin{align}
    \frac{i}{12\pi}\int_B \Tr(\Tilde{g}^{-1}\df\Tilde{g}\wedge\Tilde{g}^{-1}\df\Tilde{g}\wedge\Tilde{g}^{-1}\df\Tilde{g})
    &= \frac{i}{96\pi}\int_B i^3 \Tr(\sigma_i\sigma_j\sigma_k)\del_{\mu}\Tilde{\phi}_i\del_{\nu}\Tilde{\phi}_j\del_{\rho}\Tilde{\phi}_k \df x_{\mu}\wedge \df x_{\nu}\wedge \df x_{\rho} + \cdots\nn
    &= \frac{i}{48\pi}\int_B \del_{\mu}(\varepsilon_{ijk}\Tilde{\phi}_i\del_{\nu}\Tilde{\phi}_j\del_{\rho}\Tilde{\phi}_k) \df x_{\mu}\wedge \df x_{\nu}\wedge \df x_{\rho} + \cdots \nn 
    &= \frac{i}{48\pi}\int_{\mathbb{R}^2}\df^2x\ \varepsilon_{\mu\nu}\varepsilon_{ijk}\phi_i\del_{\nu}\phi_j\del_{\rho}\phi_k + \cdots \,,
\end{align}
where $\Tilde{\phi}_i$ is smoothly connected to $\phi_i$ on $\mathbb{R}^2$.
Compared with \Cref{eq: NLt}, we can read $\theta=1/12$. 

Note here that although the target space of $\phi_i$ is $\mathbb{R}^3$ so far, it is expected that the IR physics is dominated by its spherical components.
This is because the radius component of $\phi_i$ tends to be massive in the IR region and does not contribute to the low-energy dynamics.
This implies that the quantum fluctuation of the radius component is negligible such that $(\phi_i)^2$ is fixed to some constant (or unity by rescaling $\phi_i$) in the IR region.
Therefore, the target space of $\phi_i$ can be regarded as a two-dimensional sphere $S^2$ to describe the IR physics.
Although the decoupling of the radial mode can be intuitively understood, demonstrating it rigorously remains challenging.
Although some papers~\cite{Mitter:1988xa,Gawedzki:1986dt} have addressed this problem, there is is hard to track the change in the potential of the radial mode completely until it decouples.

The duality via non-Abelian bosonization is actually highly nontrivial. Whereas the boson theory is the WZW model whose interactions are strong and complicated, the fermion theory is a free theory.
Furthermore, studying this duality is not only attractive in particle physics, but also important in terms of condensed matter physics.
This is because the $O(3)$ nonlinear sigma model is discussed to describe the long-range behavior of antiferromagnetic spin chains, which is closely related to the Haldane conjecture~\cite{Haldane:1982rj}.

An RG analysis of the duality has already been conducted in the Witten's original paper~\cite{Witten:1983ar}, but it remains at the perturbative level.
Investigating whether this duality holds away from the fixed points, specifically in the regions beyond CFT and its perturbative vicinity, may provide deeper insights into QFTs.
Therefore, we are motivated to perform a quantitative analysis of the duality from a nonperturbative RG viewpoint.
In the following section, we apply the FRG to the $O(3)$ nonlinear sigma model with the theta term, aiming to capture the nonperturbative features of the strong interaction in the WZW model.

\section{FRG setup for \texorpdfstring{$O(3)$}{O(3)} nonlinear sigma model with theta term}
\label{sec: FRG for O(3) linear WZW model}

We study the $O(3)$ nonlinear sigma model with the theta term by using the FRG method~\cite{Wilson:1973jj} to reproduce its duality to the free fermion model nonperturbatively.
In this section, we discuss an appropriate setup for our aim.
For details of the FRG method, see the reviews~\cite{Morris:1993qb,Reuter:1993kw,Ellwanger:1993mw,Morris:1998da,Berges:2000ew,Aoki:2000wm,Bagnuls:2000ae,Polonyi:2001se,Pawlowski:2005xe,Gies:2006wv,Delamotte:2007pf,Sonoda:2007av,Igarashi:2009tj,Rosten:2010vm,Braun:2011pp}.

As mentioned in the Introduction, we face two problems in treating this model in the FRG: (i) the nonlinearity of the target space ($S^2$) and (ii) the topological nature of the theta term.
We discuss them in order in the following.

For (i), this model is a scalar field theory with the constraint $\phi_i^2 = 1$, which reflects the target space $S^2$ of the scalar fields.
Thus, the path integral for $\phi_i$ leads to nonlinear interactions.
To deal with the constraint, in the ordinary path integral analysis, an auxiliary field $\sigma(x)$ is introduced to compute the partition function as
\begin{align}
 Z=   e^{W[J]} 
 &=\int \mathcal{D}\phi\, \prod_x \delta(\phi_i(x)\phi_i(x) -1) \exp(-S[\phi]  + \int_x J_i(x)\phi_i(x))\nn
 &= \int \mathcal{D}\phi \mathcal{D}\sigma\, \exp(-S[\phi] + \int_x i\sigma(x)(\phi_i(x)\phi_i(x) -1 ) + \int_x J_i(x)\phi_i(x))\,,
\end{align}
where $S[\phi]$ is a general action of the $O(3)$ nonlinear sigma model.
However, if we naively apply this prescription to the FRG, it gives rise to another problem. The auxiliary field becomes dynamical along the RG flow through the quantum correction induced by the three-point vertex $\sigma \phi_i\phi_i$.
Consequently, theories at fixed points in this prescription could contain superfluous massless particles corresponding to $\sigma$ in addition to $\phi_i$. Those are different from the $O(3)$ nonlinear sigma model of our interest.
For this reason, we refrain from the auxiliary-field method in this work.
Although there are several works~\cite{Flore:2012ma,Flore:2012wh,Percacci:2013jpa,Efremov:2021fub} that perform FRG analyzes with the RG flow of the $O(N)$ nonlinear sigma model by solving the constraint, the truncation (approximation) in the FRG may lose information about the target space.
Therefore, we adopt the way of embedding the fields in $\mathbb{R}^3$, instead of solving the constraint.

Next, as for (ii), there is a problem peculiar to the FRG when handling topological terms $I_\mathrm{top}$, including the theta term.
In general, the topological terms reflect the global geometrical structure of the target space of the field $\phi_i$.
Therefore, it is invariant under infinitesimal variation of the field configuration without changing the boundary condition, that is, 
\begin{align}
    \delta I_\mathrm{top} = 0 \quad \text{under} \quad \phi_i(x) \to \phi_i(x) + \delta\phi_i(x) \quad \text{with} \quad \eval{\phi_i(x)}_{x\in \del M} = 0,
\end{align}
where $M$ is the spacetime manifold and $\del M$ is its boundary.
This implies that the functional derivative of the topological term always vanishes:
\begin{align}
    \fdv{I_{\mathrm{top}}}{\phi_i(x)}=0.
\end{align}
We can explicitly show this property for the theta term in the $O(3)$ nonlinear sigma model; see \Cref{sec: topological nature of theta term in O(3) nonlinear sigma}.
However, the RG flow is described in the FRG by a functional differential equation.
Therefore, it is naively expected that the contribution of these terms to the flow equations always vanishes and thus can not affect the beta function at all.

A natural way to avoid this problem is to relax the nonlinearity of the target space.
In this paper, we follow Ref.~\cite{Fukushima:2022zor}, which studies quantum mechanics for a particle on the ring $S^1\sim U(1)$ with the topological term.
Instead of dealing with $U(1)$, Ref.~\cite{Fukushima:2022zor} takes into account the radial degree of freedom to the field to extend the target space from $U(1)$ to $\mathbb{C}$.
Mimicking this way, we extend the target space $S^2$ to $\mathbb{R}^3$ in the following analysis.

Besides, we consider the one-particle irreducible (1PI) effective action $\Gamma[\phi_\text{cl}]$ in our analysis, rather than the Wilsonian effective action.
Since in the formalism of the 1PI effective action, the path integral has already been performed, the problem of the auxiliary field is expected to be resolved naturally.
Even in this approach, the 1PI effective action should satisfy some identity arising from the constraint of the target space.\footnote{
For example, let us consider two-point function in the $k\to 0$ limit.
Using the Legendre transformation, we have 
\begin{align}
    \qty(\frac{\delta^2 \Gamma[\phi]}{\delta \phi_i(x) \phi_j(y)})^{-1}
    = \eval{\frac{\delta^2 W[J]}{\delta J_i(x) J_j(y)}}_{J=J[\phi]}
    = \expval{\phi_i(x)\phi_j(y)} - \expval{\phi_i(x)}\expval{\phi_j(y)},
    \nonumber
\end{align}
where $\Gamma=\Gamma_{k=0}$ and $W[J]$ is the generating functional of connected correlation functions, defined by $W[J]\coloneqq \log Z[J]$.
If we set $y\to x,j\to i$ and take sum over $i$, we have $\expval{\phi_i^2(x)}=1$ from the constraint, and then $\Gamma$ must saturate
\begin{align}
    \qty(\frac{\delta^2 \Gamma[\phi]}{\delta \phi_i(x) \phi_i(x)})^{-1}
    = 1 - \expval{\phi_i(x)}^2.
    \nonumber
\end{align}
Note that this discussion is limited for the $k=0$ case.
For the non-zero cutoff case (i.e. within the FRG), the 1PI effective action $\Gamma_k$  should satisfy similar but modified identity reflecting on introducing the cutoff.

There are several ways to take these identities into account for our analysis.
One simple way is to modify the ansatz for the 1PI effective action so that it saturates them.
By adopting such truncation, we expect, for example, that the critical exponents would get a closer value given by CFT, since such truncation respects the information of the target space.

From the viewpoint of the flow equation, these identity gives constraint between couplings of various operators.
This yields that not all the couplings and, therefore, their flow equations are independent.
For example, if a coupling is constrained as a linear combination of other couplings ($g_0(k) = \sum_{i=1} c_i g_i(k)$), the corresponding flow equation (or the beta function) is also represented as the linear combination of those of other couplings ($\beta_{g_0} = \sum_{i=1} c_i \beta_{g_i}$).
}
In this study, we do not intend to investigate whether or not our setup maintains this constraint.
Instead, our aim is to understand the bosonization duality in viewpoint of the fixed-point structure by means of the FRG under this rough approximation.

In the following, we briefly derive the FRG equation for the 1PI effective action in two-dimensional Euclidean spacetime.
We start by defining the generating functional (Schwinger functional) $W_k[J]$ of the connected diagrams with the cutoff-function term  in two-dimensional Euclidean spacetime as 
\begin{align}
    e^{W_k[J]} = \int \mathcal{D}\phi\,
    \exp( 
     - S[\phi] - \frac{1}{2}\int_p \phi_i(-p) \mathcal{R}_k(p)\phi_i(p)
    + \int_p J_i(-p)\phi_i(p)
    )\,.
\end{align}
Here, $k$ is the IR cutoff scale and $\mathcal R_k(p)$ is the cutoff function realizing the coarse-graining process. 
In this work, we employ the Litim-type cutoff function~\cite{Litim:2001up} which is given as
\begin{align}
\mathcal R_k(p) =  Z (k^2-p^2)\theta(k^2 - p^2)\,,
\end{align}
where $Z$ is the wavefunction renormalization factor. 
Then, the cutoff-dependent 1PI effective action $\Gamma_k$ is given through the Legendre transformation~\cite{Wetterich:1992yh} such that
\begin{align}
    \Gamma_k [\phi_\text{cl}] = \sup_J\eval{\qty(-W_k[J] + \int_p J_i(p)\phi_{\text{cl},i}(p))}_{J=J_k[\phi]} - \frac{1}{2} \int_p \phi_{\text{cl},i}(-p) \mathcal{R}_k(p) \phi_{\text{cl},i}(p)\,,
    \label{eq: 1PI effective average action}
\end{align}
where the ``classical field" $\phi_{\text{cl}}$ is defined by
\begin{align}
\phi_{\text{cl},i}(p) = \langle \phi_i(p) \rangle = \frac{\delta W_k[J]}{\delta J_i}\,.
\end{align}
Hereafter, we omit ``cl" on the classical field for the sake of notational simplicity.
By taking the scale derivative for \cref{eq: 1PI effective average action}, we arrive at the flow equation (``Wetterich equation") of the 1PI effective action, which is given by
\begin{align}
\p_t \Gamma_k= \frac{1}{2} \Tr \left[ \left( \Gamma_k^{(2)} + \mathcal R_k \right)^{-1} \p_t \mathcal R_k \right]\,.
\label{eq: Wetterich equation}
\end{align}
where $\p_t=k\p_k$ is the derivative with respect to the dimensionless scale $t=\log k$ and $\Tr$ denotes functional trace over internal indices $(i,j,\ldots)$ and the momentum $(p,q,\ldots)$.
Here, $\Gamma^{(2)}_k$ is is the full two-point function (inverse propagator) obtained by the second-order functional derivative with respect to the fields $\phi_i$, i.e.,
\begin{align}
 \Big(   \Gamma^{(2)}_k(p,q) \Big)_{ij} = \sfdv{\Gamma_k}{\phi_i(-p)}{\phi_j(q)}\,.
\end{align}
The equation~\eqref{eq: Wetterich equation} describes the change in the scale-dependent 1PI effective action $\Gamma_k$ along $k$.

The 1PI effective action $\Gamma_k$ contains an infinite number of effective operators in principle; however, in practice, we have to truncate it so as to get a set of RG flow equations for the couplings in closed form.
In this work, we make the following truncation ansatz, which is composed of up to four vertex functions:\footnote{
As previously discussed, we extend the target space from the 2-sphere $S^2$ to the three-dimensional Euclidean space $\mathbb{R}^3$.
Whereas we expect that the extended theory converges to the desired RG trajectory of the $O(3)$ non-linear sigma model in the continuum limit, it is a subtle challenge to confirm this expectation.
In the case of a finite volume, the field configuration with the linearized target space contributes to the partition function with the weight $\Lambda^2 \Omega e^{-g/\Lambda^2}$, where $\Omega$ denotes the spacetime volume, $g$ represents the quartic coupling and $\Lambda$ is a UV cutoff.
While it might seem that the continuum $O(3)$ non-linear sigma model is recovered as $\Lambda$, $\Omega$, and $g$ tend to infinity, an obtained continuum theory is highly sensitive to how these limits are taken.
}
\begin{multline}
\label{eq: anstaz for 1PI effective action}
    \Gamma_k = \frac{1}{2} \int_p (Zp^2+m^2)\phi_i(p)\phi_i(-p)
    + \int_{\{p,q,r\}} i V_3(p,q,r) \varepsilon_{ijk} \varepsilon_{\mu\nu} p_{\mu}q_{\nu} \phi_i(p) \phi_j(q) \phi_k(r)
\\
    + \int_{\{p,q,r,s\}} V_4(p,q,r,s) \phi_i(p)\phi_i(q) \phi_j(r) \phi_j(s),
\end{multline}
where we have used a short-hand notation
\begin{align}
    &\int_p=\int\frac{\df^2p}{(2\pi)^2},& 
    &\int_{\{p_1,\ldots,p_n\}} = \int \frac{\df^2 p_1}{(2\pi)^2}\cdots \frac{\df^2 p_n}{(2\pi)^2} \,\delta^{(4)}\qty(\sum_{i=1}^n p_i).
\end{align}
The vertex functions $V_3$ and $V_4$ are functions of the external momenta.
Although we can derive the flow equations for $V_3$ and $V_4$ by keeping general momentum configurations, practically solving these equations becomes very costly in numerical calculations.
We are interested in the effective action in the low-energy region, where the momentum effect would be suppressed.
Therefore, we expand these vertex functions in terms of external momenta, corresponding to the derivative expansion and truncate terms up to the second-order derivatives in each vertex:\footnote{
Note that including the derivative terms in our truncation is crucial in this setup. Without these terms, it is impossible to obtain the fixed point corresponding to the WZW model. Once the derivative terms are introduced and the appropriate fixed point is identified, the critical exponents can be further refined by extending the theoretical space, or equivalently, by incorporating higher-order interactions.
}
\begin{align}
\label{eq: anstaz for vertices}
    V_3(p,q,r) &= \frac{\theta}{6}, \\
    V_4(p,q,r,s) &= \frac{1}{24}
    \qty(\lambda_3 -\frac{\lambda_1}{2}(p \cdot q + r\cdot s) - \frac{\lambda_2}{4}(p+q)\cdot (r+s)).
\end{align}
This parametrization of the vertex functions is different from that of \Cref{eq: NLt}.
It is just for computational convenience and does not affect any physical results.
In particular, the scaling dimensions, calculated in the next section, are independent of the arbitrariness of parametrization.

\section{RG flow, fixed point and scaling dimensions}
\label{sec: RG flow and fixed points}

In this section, we calculate the RG flow equations for couplings under the truncation of \cref{eq: anstaz for 1PI effective action}.
After that, we search fixed points of these RG flow equations and calculate the scaling dimensions around the fixed points.
We also examine the consistency with the duality between the $O(3)$ nonlinear sigma model with the theta term and the free fermion model with two-flavor.

\subsection{Flow equations}
Let us start by calculating the RG flow equations for the couplings.
Substituting the ansatz \eqref{eq: anstaz for 1PI effective action} and each vertex \eqref{eq: anstaz for vertices} into the Wetterich equation~\eqref{eq: Wetterich equation}, we obtain a set of the flow equations
\begin{align}
\label{eq: RG flow equation}
    \del_t \Tilde{g}_i = \beta_{\Tilde{g}_i} (\Tilde{g}),
\end{align}
where $\Tilde{g}_i=(\Tilde{m}^2,\Tilde{\theta},\Tilde{\lambda}_1,\Tilde{\lambda}_2,\Tilde{\lambda}_3)$ are the dimensionless renormalized couplings defined as
\begin{align}
    \Tilde{m}^2 \coloneqq Z^{-1}k^{-2}m^2, \quad
    \Tilde{\theta} \coloneqq Z^{-\frac{3}{2}}\theta, \quad 
    \Tilde{\lambda}_1 \coloneqq Z^{-2}\lambda_1,\quad
    \Tilde{\lambda}_2 \coloneqq Z^{-2}\lambda_2,\quad
    \Tilde{\lambda}_3 \coloneqq Z^{-2}k^{-2}\lambda_3,
\end{align}
and $\beta_{\tilde g_i}$ denotes the beta function for each coupling.
The explicit form of the beta functions is obtained as
\begin{subequations}
\label{eq: beta functions}
\begin{align}
    \beta_{\tilde m^2} &= -(2-\eta) \Tilde{m}^2 - \frac{3 \Tilde{\lambda}_1+\Tilde{\lambda}_2 +20 \Tilde{\lambda}_3}{48 \pi (1+\Tilde{m}^2)^2},\\
    \beta_{\tilde \theta} &= \frac{3}{2}
    \eta \Tilde{\theta} 
    + \frac{ \Tilde{\theta} (8 \Tilde{\lambda}_1 - 3\Tilde{\lambda}_2 )
    }{
    32\pi (1 + \Tilde{m}^2 ) ^3
    },\\
    \beta_{\Tilde{\lambda}_3} & = -(2-2\eta)\Tilde{\lambda}_3
    + \frac{
      3\Tilde{\lambda}_1^2
     + \Tilde{\lambda}_2^2
     + 18 \Tilde{\lambda}_2 \Tilde{\lambda}_3
     + 132 \Tilde{\lambda}_3^2
     + 2 \Tilde{\lambda}_1 \Tilde{\lambda}_2
     + 30\Tilde{\lambda}_1 \Tilde{\lambda}_3
     }{72\pi(1+\Tilde{m}^2)^3},\\
     \beta_{\Tilde{\lambda}_1} &= 2 \eta  \Tilde{\lambda}_1 + \frac{
     3\Tilde{\lambda}_1^2
     +\Tilde{\lambda}_2^2
     +12\Tilde{\lambda}_2\Tilde{\lambda}_3
     +2\Tilde{\lambda}_1\Tilde{\lambda}_2
     +20\Tilde{\lambda}_1\Tilde{\lambda}_3
     }{24\pi(1+\Tilde{m}^2)^3}
    - \frac{
    2\Tilde{\lambda}_3^2 - 36\Tilde{\lambda}_3\Tilde{\theta}^2
    }{3\pi(1+\Tilde{m}^2)^4},\\
    \beta_{\Tilde{\lambda}_2} &= 2 \eta  \Tilde{\lambda}_2
    + \frac{
      7\Tilde{\lambda}_1^2
    + 5\Tilde{\lambda}_2^2 
    + 64\Tilde{\lambda}_2\Tilde{\lambda}_3 
    + 14\Tilde{\lambda}_1\Tilde{\lambda}_2 
    + 48\Tilde{\lambda}_1\Tilde{\lambda}_3
    }{24\pi(1+\Tilde{m}^2)^3}
    - \frac{
    3\Tilde{\lambda}_3^2 + 2\Tilde{\lambda}_3\Tilde{\theta}^2
    }{\pi(1+\Tilde{m}^2)^4},
\end{align}
\end{subequations}
where $\eta$ is the anomalous dimension arising from the scalar field renormalization factor, given by
\begin{align}
    \eta \coloneqq -\frac{1}{Z}\p_t Z 
    = \frac{1}{2} \left(\frac{3 \Tilde{\lambda}_1+\Tilde{\lambda}_2}{12\pi  (1+\Tilde{m}^2)^2}+\frac{3 \Tilde{\theta} ^2}{4\pi(1+\Tilde{m}^2)^3}\right).
    \label{eq: anomalous dimension of scalar field}
\end{align}
In the beta functions, the first terms correspond to the canonical scaling effects which reflect the mass dimension of each coupling, combined with the anomalous dimension from the field renormalization.

\subsection{Fixed point structure}
Let us explore the fixed points $g_i^* = (\Tilde{m}^{2*},\theta^*,\lambda_1^*,\lambda_2^*,\lambda_3^*)$ in the above RG flow equation~\eqref{eq: RG flow equation}, which are defined by $\beta_{g_i}(g^*)=0$.
The trivial solution is the Gaussian fixed point where $\tilde g_i^*=0$ for all $i$.
By numerical calculation, we search for several nontrivial fixed points where $\tilde g_i^*\neq 0$.
In this work, we are interested in the case of the nonzero real value for $\theta$.
We found only one such nontrivial fixed point.
The fixed point values of the couplings, as well as the Gaussian fixed point, are summarized in \Cref{tab:Fixed Points}.
\begin{table}[t]
    \centering
\begin{tabular}{lcccccc}
\toprule
   Fixed Point (FP) & $\Tilde{m}^2_*$ & $\Tilde{\theta}_*$ & $\Tilde{\lambda}_{1*}$ & $\Tilde{\lambda}_{2*}$ & $\Tilde{\lambda}_{3*}$ & $\eta^*$\\
\midrule
(1) Gaussian FP & 0 & 0 & 0 & 0 & 0 & 0 \\[1ex]
(2) Nontrivial FP & $-0.166948$ & $\pm0.0961412$ & 0.168737 & 1.76901 & 1.59369 & 0.0453913 \\
\bottomrule
\end{tabular}
    \caption{The fixed point values of the couplings and the anomalous dimension: (1) the Gaussian fixed point and (2) the nontrivial fixed point. The latter is expected to be the fixed point corresponding to $O(3)$ nonlinear sigma model with the theta term.}
    \label{tab:Fixed Points}
\end{table}

Note that at the Gaussian fixed point, where
\begin{align}
\Tilde{m}^{2*} =
\Tilde{\theta}^* = \Tilde{\lambda}_1^* = \Tilde{\lambda}_2^* = \Tilde{\lambda}_3^* = 0,
\label{eq: Gaussian fixed point}
\end{align}
the anomalous dimension of the field vanishes:
\begin{align}
    \eta^* = 0.
\end{align}
On the other hand, at the nontrivial fixed point, where
\begin{align}
  \Tilde{m}^{2*} = -0.166948, \quad 
  \Tilde{\theta}^* = \pm 0.0961412, \quad 
  \Tilde{\lambda}_1^* = 0.168737, \quad
  \Tilde{\lambda}_2^* = 1.76901, \quad
  \Tilde{\lambda}_3^* = 1.59369,
  \label{eq: nontrivial fixed point}
\end{align}
the anomalous dimension of the field takes the nontrivial value:
\begin{align}
    \eta^* = 0.0453913.
\end{align}
It should be noted here that the plus and minus signs of $\tilde\theta^*$ reflect $Z_2$ symmetry of the scalar fields ($\phi_i \to -\phi_i$) for which the flow  equation~\eqref{eq: Wetterich equation} is invariant.
The nontrivial fixed point \eqref{eq: nontrivial fixed point} is of our interest and is studied in detail in the next subsection.

Here, we comment on the value of the coupling at the non-trivial fixed point~\eqref{eq: nontrivial fixed point}.
Although this fixed point is expected to correspond to the critical WZW model, the values of the couplings can be different from those in the continuum theory due to the introduction of the explicit cutoff.
In addition, the present ansatz~\eqref{eq: anstaz for 1PI effective action} spoils the $2\pi$-periodicity of the topological coupling.
For these reasons, we do not take the value of the coupling itself seriously, while we focus on the critical exponents in this paper.

\begin{table}[t]
    \centering
\begin{tabular}{lcccccc}
\toprule
 Eigenvalue & $\vartheta_i$ & 2 & 0 & 0 & 0 & 2\\
\midrule
 Eigenvector 
 & $\mqty(\Tilde{m}^2 \\ \Tilde{\theta} \\ \Tilde{\lambda}_1 \\ \Tilde{\lambda}_2 \\ \Tilde{\lambda}_3)$
 & $\mqty(1 \\ 0 \\ 0 \\ 0 \\ 0)$ 
 & $\mqty(0 \\ 1 \\ 0 \\ 0 \\ 0)$ 
 & $\mqty(0 \\ 0 \\ 1 \\ 0 \\ 0)$ 
 & $\mqty(0 \\ 0 \\ 0 \\ 1 \\ 0)$ 
 & $\mqty(0 \\ 0 \\ 0 \\ 0 \\ 1)$ \\
\bottomrule 
\end{tabular}
    \caption{Eigensystem at the Gaussian fixed point \eqref{eq: Gaussian fixed point}. From the left to the right, $i=1,2,\cdots,5$.}
    \label{tab:Eigensystem at Gaussian}
\end{table}

\subsection{Scaling dimensions}
Let us study the RG flow structure around the fixed points exhibited in \Cref{tab:Fixed Points} by calculating the scaling dimensions.
To this end, we consider a slight perturbation of the couplings around these fixed points such that $\Tilde{g}=\Tilde{g}_*+\delta \Tilde{g}$.
In this case, the RG flow equations~\eqref{eq: RG flow equation} read
\begin{align}
    \del_t (\Tilde{g}^* + \delta \Tilde{g})_i = \beta_{\Tilde{g}_i}(\Tilde{g}^* + \delta \Tilde{g}) = -{\mathcal T}_{ij}\delta \Tilde{g}_j + \mathcal{O}\qty((\delta \Tilde{g})^2).
\end{align}
where we defined the stability matrix $\mathcal{T}_{ij}$ as 
\begin{align}
{\mathcal T}_{ij} \coloneqq -\eval{\frac{\del \beta_{\Tilde{g}_i}}{\del \Tilde{g}_j}}_{\Tilde{g}^*}.
\end{align}
Here, we have performed the Taylor expansion for the beta functions in the deviation $\delta \tilde g_i$ and have taken up to its linear order. Note that $\p_t \tilde g_i^*=0$ and the lowest order of the Taylor expansion vanishes by the definition of the fixed point, i.e., $\beta_{\Tilde{g_i}}(\tilde g^*)=0$. We obtain the linearized RG flow equations for the deviation $\delta \Tilde{g}_i$ as
\begin{align}
    \del_t \delta \Tilde{g}_i = -{\mathcal T}_{ij}\delta \Tilde{g}_j.
\end{align}
The eigenvalues of the stability matrix at a fixed point, i.e.,
\begin{align}
\vartheta_i = \text{eig} \left( {\mathcal T}_{ij} \right),
\label{eq: eigenvalues of stability matrix}
\end{align}
are called the scaling dimensions or the critical exponents.
The corresponding eigenvectors define the linear combinations of the operator basis.
Such linearly combined operators are called the scaling operators.
A scaling operator with positive/negative scaling dimension is called a relevant/irrelevant operator.
When we perturb a fixed point theory in the direction of relevant/irrelevant operators, it goes away from/converges to the fixed point along the RG flow.
If a scaling dimension is zero, the corresponding scaling operator is called marginal.

When the off-diagonal parts of the stability matrix is negligible, the scaling dimension for a coupling approximately reads
\begin{align}
\vartheta_i \simeq \text{(canonical dimension)} - \text{(anomalous dimension)}.
\label{eq: naive critical exponent}
\end{align}
Thus, the scaling dimensions obtains contributions from the canonical and anomalous dimensions.
Note here that ``anomalous dimension" in \Cref{eq: naive critical exponent} implies not only from the scalar field renormalization factor \eqref{eq: anomalous dimension of scalar field}, but also from loop effects at a fixed point $\tilde g_i^*$ in the beta functions.
In the following, we compute the scaling dimensions for each fixed point and discuss the behavior of the flow of the couplings.

\paragraph{Gaussian fixed point}
First, let us briefly discuss the Gaussian fixed point~\eqref{eq: Gaussian fixed point}. We summarize the eigenvalues (scaling dimensions) and corresponding eigenvectors of the stability matrix \eqref{eq: eigenvalues of stability matrix} in Table~\ref{tab:Eigensystem at Gaussian}. 
The system at the Gaussian fixed point becomes massless and has no interaction.
Because a free theory induces no anomalous dimension, the scaling dimensions around this fixed point are identical to the canonical scaling.
Accordingly, the scaling operators are also identical to the operators corresponding to the original couplings.
Note that the scaling dimension of the quartic operator without derivatives (momenta), corresponding to $\lambda_3$, is two, so that this operator is relevant at the Gaussian fixed point, as well as the mass term.

\begin{table}[t]
    \centering
\begin{tabular}{lccccc}
\toprule
 Eigenvalue & $\vartheta_i$ &  $1.8$ & $0.092$ & $-2.9 \pm 1.1i$ & $-0.50$ \\
\midrule 
 Eigenvector 
 & $\mqty(\Tilde{m}^2 \\ \Tilde{\theta} \\ \Tilde{\lambda}_1 \\ \Tilde{\lambda}_2 \\ \Tilde{\lambda}_3)$
 & $\mqty(1.00 \\ 0.0050 \\ 0.0034 \\ -0.0013 \\ 0.056)$ 
 & $\mqty(-0.011 \\ 1.0 \\ 0.039 \\ -0.0077 \\ 0.00038)$ 
 & $\mqty(0.94 \\-0.016 \pm 0.071i \\0.0010 \mp 0.10i \\0.046 \mp 0.087i \\-0.31 \pm 0.034i)$ 
 & $\mqty(-0.0092 \\ -0.98 \\ 0.20 \\ -0.042 \\ 0.000035)$ 
 \\ 
\bottomrule 
\end{tabular}
    \caption{Eigensystem at the nontrivial fixed-point \eqref{eq: nontrivial fixed point}. From the left to the right, $i=1,2,\cdots,5$.}
    \label{tab:Eigensystem at nontrivial}
\end{table}

\paragraph{Nontrivial fixed point}
We turn to the discussion on the RG flow structure at the nontrivial fixed point~\eqref{eq: nontrivial fixed point}.
We summarize the eigensystem of the stability matrix at this fixed point in \Cref{tab:Eigensystem at nontrivial}.
First of all, it can be seen from Table~\ref{tab:Eigensystem at nontrivial} that the scaling dimensions are totally different from those of the Gaussian fixed point. 
This indicates that at the nontrivial fixed point, large anomalous dimensions for the couplings are induced by the quantum effect, as mentioned in \Cref{eq: naive critical exponent}.
Accordingly, the eigenvectors have nontrivial components, indicating mixing between the operators; however, we cannot identify which scaling dimension corresponds to the original operator.
To roughly grasp the correspondence between the scaling operators and the original ones, it is helpful to consider only the diagonal parts of the stability matrix.
Supposing that the off-diagonal parts of the stability matrix take small values, we approximately read off
\begin{align}
\vartheta_i \approx \mathcal{T}_{ii}.
\end{align}
At the nontrivial fixed point, the right-hand side of this equation is given by
\begin{align}
&\mathcal{T}_{11} = 1.2,&
&\mathcal{T}_{22} = 0.0076,&
&\mathcal{T}_{33} = -0.95,&
&\mathcal{T}_{44} = -3.0,&
&\mathcal{T}_{55} = -1.6.
\label{eq:approximated scaling dimensions}
\end{align}
From these values and \Cref{tab:Eigensystem at nontrivial}, we can see that 
\begin{align}
&\vartheta_1 \simeq \mathcal{T}_{11},& 
&\vartheta_2 \simeq \mathcal{T}_{22}.
\end{align}
Therefore, the scaling operators corresponding to $\vartheta_1$ and $\vartheta_2$ are roughly identified with the mass and theta terms, respectively.

On the other hand, it is difficult to identify the scaling dimensions for the irrelevant operators.
We notice here that there are complex scaling dimensions as $\vartheta_{3,4}=-2.9\pm 1.1i$, whereas the values of $\mathcal{T}_{33}$, $\mathcal{T}_{44}$ and $\mathcal{T}_{55}$ are real. 
The existence of an imaginary part leads to the fact that the RG flow of corresponding couplings shows rotational behavior around the nontrivial fixed point. 
This implies that the corresponding scaling operators are mixed by interactions.\footnote{
Complex values of scaling dimensions are often observed in the analysis of asymptotically safe quantum gravity.
See, e.g., Refs.~\cite{Reuter:1996cp,Souma:1999at,Oda:2015sma,Hamada:2017rvn}.
}
Because its real part is negative, these flows are irrelevant and thus converge to the fixed point while rotating.
At this point, we cannot conclude whether or not the appearance of the imaginary part in the scaling dimensions is due to a somewhat crude truncation employed in the present setup for the effective action. 
Hence, this issue may be resolved with a more refined truncation.

From the identification $\vartheta_1$ with $\mathcal{T}_{11}$, we observe a decrease in the scaling dimension of the mass term from the canonical mass dimension of two because of mixing with the other interactions.
This is consistent with the non-Abelian bosonization.
From this duality, this critical exponent is supposed to be $3/2$ as explained below.
Firstly, consider the mass operator \eqref{eq: duality of mass term} with $SU(2)$-valued field $g$, but not with $U(2)$-valued field.
It is calculated with \Cref{eq: SU(2) field in terms of O(3) vector} as
\begin{align}
    \tr (g+g^\dag)&=\tr \left[\exp(i\frac{\sigma_i}{2}\phi_i) + \exp(-i\frac{\sigma_i}{2}\phi_i) \right] \nn
     &= (\text{const.}) - \frac{1}{4} \tr(\sigma_i\sigma_j) \phi_i\phi_j + \frac{1}{192} \tr(\sigma_i\sigma_j\sigma_k\sigma_l) \phi_i \phi_j \phi_k \phi_l + \order{\phi^6} \nn
     &= (\text{const.}) - \frac{1}{2}\phi_i\phi_i + \frac{1}{96} \phi_i \phi_i \phi_j \phi_j + \order{\phi^6}.
     \label{eq: composite fermions}
\end{align}
Thus, naively, this operator corresponds to the relevant operator (the mass term mixed with the quartic term) around the non-trivial fixed point.
In addition, we have to take the $U(1)$ part of $U(2)$ into account.
From \Cref{eq: decompose U(N) to U(1) and SU(N)}, 
there is an overall factor $e^{\pm i\varphi}$ in front of $g$ and $g^\dag$, whose operator dimension is given by 1/2 \footnote{
This can be calculated using conformal field theory.
See \cite{DiFrancesco:1997nk} for example.
}.
  On the other hand, the operator dimension of the mass term in the two-flavor free fermion theory is unity, as well as that of $\tr(g+g^\dag)$ in $U(2)$ WZW model.
Due to the existence of $U(1)$ part, the operator dimension of Eq.~\eqref{eq: composite fermions} is less than that for $U(2)$ by 1/2. 
Therefore, the critical exponent of Eq.~\eqref{eq: composite fermions} is 3/2, which can be obtained by subtracting the operator dimension (1/2) from the spacetime dimension (2)\footnote{This is also consistent with the direct analysis of $SU(2)$ WZW model at level-1 and a compact free scalar theory dual to it. See \cite{DiFrancesco:1997nk} for example.}.

Next, we discuss the value of $\vartheta_2$, which is approximately identified with $\mathcal{T}_{22}$.
It takes a positive small value, and this result may reflect the topological nature of the theta term. 
In continuum theories, the coefficients of topological terms are discretized because they represent the winding number associated with global geometrical structures of field configurations. 
Thus, they should not be affected by continuous variations, such as the renormalization group transformations.
Although the scaling dimension of the theta term should vanish from this discussion, some small effects on it would be induced by the renormalization procedure in the FRG, which requires explicit cutoff regularization.

Finally, the values of $\vartheta_3,\vartheta_4,\vartheta_5$ tell us that all the four-point interactions at the nontrivial fixed points, in contrast to those at the Gaussian fixed point, tend to be irrelevant by mixing with the other interactions.
This is also consistent with non-Abelian bosonization, where higher-point vertices such as $(\Bar{\psi}_i\psi_i)^2$ and $(\Bar{\psi}_i\gamma_{\mu}\psi_i)^2$ are marginal or irrelevant in the fermion model.

\section{Conclusions and Discussions}
\label{sec: conclusion and discussion}
In this paper, we have studied the $O(3)$ nonlinear sigma model with the theta term within the FRG, aiming to derive its duality between the free fermion model with two-flavor and the $U(2)$ WZW model nonperturbatively.

When we analyze this model with the FRG, there are two problems that must be taken into account: the nonlinearity of the target space $S^2$ and the topological nature of the theta term.
To address these problems, we employed the 1PI effective action and relaxed the nonlinearity of the target space from $S^2$ to $\mathbb{R}^3$.
There, we truncated the 1PI effective action with up to four-point vertices and up to a second-order derivative for each vertex with preserving the $O(3)$ symmetry.

Within this truncation, we calculated the RG flow equation for couplings using the Wetterich equation with the Litim-type cutoff function.
We found a nontrivial fixed point in addition to the Gaussian fixed point, and studied the RG flow structure around them.
In the Gaussian fixed point, there are two relevant operators (mass term and four-point vertex) and three marginal operators (the theta term and two types of four-point vertices with second-order derivative).
On the other hand, around the nontrivial fixed point, this situation drastically changes such that there are two relevant operators, which are approximately identified with the mass term and the theta term, and three irrelevant operators.
It is noteworthy that the four-point vertices become irrelevant, although those are relevant around the Gaussian fixed point.
We expect that the $O(3)$ nonlinear sigma model at this nontrivial fixed point corresponds to the $SU(2)$ WZW model or the free fermion model.

The first relevant operator has the scaling dimension $\vartheta_1=1.8$ and the corresponding operator is approximately a linear combination of the mass term and the four-point vertex without derivative.
This is consistent with the dictionary of non-Abelian bosonization duality.
Note that the canonical scaling dimension of the mass term in the $O(3)$ nonlinear sigma model is two.
The equivalence to the free fermion theory implies that the scaling dimension at the nontrivial fixed point should be $3/2$, while our calculation yields a little bit larger value as $\vartheta_1=1.8$.
This discrepancy is due to the fact that, as can be seen from \Cref{eq: composite fermions}, the mass operator of the fermion fields involves an infinite number of scalar-field interactions.

The second relevant operator has a scaling dimension $\vartheta_2=0.092$ (or $\mathcal{T}_{22}=0.0076$ in the approximated case) and the corresponding operator is approximately identified with the theta term.
Since its scaling dimension is relatively smaller than that of the first relevant operator, this fact is consistent with the discussions of nonrenormalization of coupling for topological terms, i.e., the topological interactions are exactly marginal.

We expect that these critical exponents $\vartheta_1$ and $\vartheta_2$ tend to converge to $3/2$ and zero, respectively, by increasing the truncation level.
It is noteworthy that the FRG calculation in the present work captures well the fundamental features of the duality between the different models, despite the inclusion of only the five couplings and the difference of their target spaces.

Finally, let us make several comments on future prospects. 
First, our result is insufficient to prove the non-Abelian bosonization duality, in the sense that the scaling dimension corresponding to the relevant operator (the mass term) is bigger than 3/2, as expected by the duality to the free fermion model.
This discrepancy will be resolved by improving our truncation.

Second, in order to make our analysis more robust, it is worthwhile to compare values of the central charge between these dual models with the FRG.
The Gaussian fixed point corresponds to the free $O(3)$ vector boson theory, while the nontrivial fixed point is expected to correspond to the two-flavor free fermion model.
The central charge is three for the former and unity for the latter.
The verification of these values is crucial for proving the non-Abelian bosonization duality.
The RG flow equation for the $c$-function, as derived in Ref.~\cite{Codello:2013iqa}, may be available for this purpose.

Another direction is to analyze the Wilsonian effective action with an auxiliary field, which is not employed in this work.
To this end, we may utilize Gradient Flow Exact Renormalization Group~\cite{Sonoda:2020vut,Makino:2018rys,Sonoda:2019ibh,Miyakawa:2021hcx,Miyakawa:2021wus,Sonoda:2022fmk,Miyakawa:2023yob,Abe:2022smm,Haruna:2023spq}, which is a recently proposed framework to define an RG flow based on diffusion equations.
In this framework, we can define an RG flow that preserves the constraint ($\phi_i \phi_i =1$) of the fields by choosing a diffusion equation tailored for $O(N)$ nonlinear sigma models~\cite{Makino:2014sta,Makino:2014cxa,Aoki:2014dxa}.
The dynamical hadoronization~\cite{Gies:2001nw,Fukushima:2021ctq,Fukushima:2023wnl} in the FRG might also provide an alternative way for that purpose.
This method can capture information on composite states by introducing cutoff-dependent auxiliary fields.
Utilizing these frameworks, we may discuss the validity of relaxing the nonlinearity of the target space of the fields and topological nature of the theta term from another viewpoint.
We leave these studies to future work.

\subsection*{Acknowledgments}
We thank Hiroshi Suzuki and Yuya Tanizaki for helpful discussions.
The work of M.\ Y. is supported by the National Science Foundation of China (NSFC) under Grant No.~12205116 and the Seeds Funding of Jilin University.

\appendix
\section{Topological nature of theta term in \texorpdfstring{$O(3)$}{O(3)} nonlinear sigma}
\label{sec: topological nature of theta term in O(3) nonlinear sigma}
In this section, we show the topological nature of the theta term in the $O(3)$ nonlinear model, defined as
\begin{align}
    I_{\mathrm{theta}} = \frac{i}{4\pi}\int_x \epsilon_{ijk} \epsilon_{\mu\nu} \phi_i(x) \del_\mu \phi_j(x) \del_\nu \phi_k(x)
\end{align}
If we consider its variation, we get
\begin{align}
    \delta I_{\mathrm{theta}} = \frac{3i}{4\pi}\int_x \epsilon_{ijk} \epsilon_{\mu\nu} \delta\phi_i(x) \del_\mu \phi_j(x) \del_\nu \phi_k(x)
     = \frac{3i}{2\pi}\int_x \epsilon_{ijk} \delta\phi_i(x) \del_1 \phi_j(x) \del_2 \phi_k(x).
     \label{eq: variation of WZ term}
\end{align}
Here, the variation must be taken to keep the target space invariant:
\begin{align}
    0 = \delta(\phi_i(x)^2) = \phi_i(x) \delta \phi_i(x).
    \label{eq: orthogonality of phi and delta phi}
\end{align}
In addition, the derivative of the field $\phi_i(x)$ is orthogonal to the vector $\phi_i(x)$:
\begin{align}
    0 = \del_\mu (\phi_i(x)^2) = \phi_i(x) \del_\mu \phi_i(x).
    \label{eq: orthogonality of phi and del phi}
\end{align}
Let us consider what becomes the coefficient vector of $\delta\phi_i(x)$ in \Cref{eq: variation of WZ term}.
Here, the points are that the field $\phi_i(x)$ is embedded in $\mathbb{R}^3$, whose dimension is three, 
that $\del_1\phi_i(x)$ and $\del_2\phi_i(x)$ are not parallel in general, and that $\phi_i(x)$ is orthogonal to both $\del_1\phi_i(x)$ and $\del_2 \phi_i(x)$ as can be seen from \Cref{eq: orthogonality of phi and del phi}.
Taking these facts into accounts, it is deduced that $\epsilon_{ijk} \del_1\phi_j(x) \del_2\phi_k(x)$ is parallel to $\phi_i(x)$:
\begin{align}
    \epsilon_{ijk}\, \del_1\phi_j(x) \del_2\phi_k(x) = c \phi_i(x),
\end{align}
with some $x$-independent constant $c$.
Note that $\del_\mu\phi_i(x)\, (\mu=1,2)$ is orthogonal to this operator:
\begin{align}
    \del_1 \phi_i\cdot \epsilon_{ijk} \del_1\phi_j(x) \del_2\phi_k(x) 
    = \del_2 \phi_i\cdot \epsilon_{ijk} \del_1\phi_j(x) \del_2\phi_k(x) 
    = 0.
\end{align}
Using this expression and the orthogonality of $\delta \phi_i(x)$ and $\phi_i(x)$ from \Cref{eq: orthogonality of phi and delta phi}, we find that the variation of the theta term is calculated as
\begin{align}
    \delta I_{\mathrm{theta}} = \frac{3i}{2\pi}
    \int_x \delta \phi_i(x) \cdot c\phi_i(x) = 0.
\end{align}
Then, we conclude that the functional derivative of the theta term always vanishes:
\begin{align}
    \fdv{I_{\mathrm{theta}}}{\phi_i} = 0.
\end{align}

\bibliographystyle{JHEP} 
\bibliography{ref}

\providecommand{\href}[2]{#2}\begingroup\raggedright\begin{thebibliography}{10}

\bibitem{Maldacena:1997re}
J.~M. Maldacena, \emph{{The Large N limit of superconformal field theories and
  supergravity}}, \href{https://doi.org/10.4310/ATMP.1998.v2.n2.a1}{\emph{Adv.
  Theor. Math. Phys.} {\bfseries 2} (1998) 231}
  [\href{https://arxiv.org/abs/hep-th/9711200}{{\ttfamily hep-th/9711200}}].

\bibitem{Witten:1998qj}
E.~Witten, \emph{{Anti-de Sitter space and holography}},
  \href{https://doi.org/10.4310/ATMP.1998.v2.n2.a2}{\emph{Adv. Theor. Math.
  Phys.} {\bfseries 2} (1998) 253}
  [\href{https://arxiv.org/abs/hep-th/9802150}{{\ttfamily hep-th/9802150}}].

\bibitem{Coleman:1974bu}
S.~R. Coleman, \emph{{The Quantum Sine-Gordon Equation as the Massive Thirring
  Model}}, \href{https://doi.org/10.1103/PhysRevD.11.2088}{\emph{Phys. Rev. D}
  {\bfseries 11} (1975) 2088}.

\bibitem{Witten:1983ar}
E.~Witten, \emph{{Nonabelian Bosonization in Two-Dimensions}},
  \href{https://doi.org/10.1007/BF01215276}{\emph{Commun. Math. Phys.}
  {\bfseries 92} (1984) 455}.

\bibitem{Senechal:1999us}
D.~Senechal, \emph{{An Introduction to bosonization}},  in \emph{{CRM Workshop
  on Theoretical Methods for Strongly Correlated Fermions}}, 8, 1999,
  \href{https://arxiv.org/abs/cond-mat/9908262}{{\ttfamily cond-mat/9908262}}.

\bibitem{Haldane:1982rj}
F.~D.~M. Haldane, \emph{{Continuum dynamics of the 1-D Heisenberg
  antiferromagnetic identification with the O(3) nonlinear sigma model}},
  \href{https://doi.org/10.1016/0375-9601(83)90631-X}{\emph{Phys. Lett. A}
  {\bfseries 93} (1983) 464}.

\bibitem{Haldane:1983ru}
F.~D.~M. Haldane, \emph{{Nonlinear field theory of large spin Heisenberg
  antiferromagnets. Semiclassically quantized solitons of the one-dimensional
  easy Axis Neel state}},
  \href{https://doi.org/10.1103/PhysRevLett.50.1153}{\emph{Phys. Rev. Lett.}
  {\bfseries 50} (1983) 1153}.

\bibitem{Wilson:1971bg}
K.~G. Wilson, \emph{{Renormalization group and critical phenomena. 1.
  Renormalization group and the Kadanoff scaling picture}},
  \href{https://doi.org/10.1103/PhysRevB.4.3174}{\emph{Phys. Rev. B} {\bfseries
  4} (1971) 3174}.

\bibitem{Wilson:1973jj}
K.~G. Wilson and J.~B. Kogut, \emph{{The Renormalization group and the epsilon
  expansion}}, \href{https://doi.org/10.1016/0370-1573(74)90023-4}{\emph{Phys.
  Rept.} {\bfseries 12} (1974) 75}.

\bibitem{Wegner:1972ih}
F.~J. Wegner and A.~Houghton, \emph{{Renormalization group equation for
  critical phenomena}},
  \href{https://doi.org/10.1103/PhysRevA.8.401}{\emph{Phys. Rev. A} {\bfseries
  8} (1973) 401}.

\bibitem{Polchinski:1983gv}
J.~Polchinski, \emph{{Renormalization and Effective Lagrangians}},
  \href{https://doi.org/10.1016/0550-3213(84)90287-6}{\emph{Nucl. Phys. B}
  {\bfseries 231} (1984) 269}.

\bibitem{Wetterich:1992yh}
C.~Wetterich, \emph{{Exact evolution equation for the effective potential}},
  \href{https://doi.org/10.1016/0370-2693(93)90726-X}{\emph{Phys. Lett.}
  {\bfseries B301} (1993) 90}
  [\href{https://arxiv.org/abs/1710.05815}{{\ttfamily 1710.05815}}].

\bibitem{Morris:1993qb}
T.~R. Morris, \emph{{The Exact renormalization group and approximate
  solutions}}, \href{https://doi.org/10.1142/S0217751X94000972}{\emph{Int. J.
  Mod. Phys.} {\bfseries A9} (1994) 2411}
  [\href{https://arxiv.org/abs/hep-ph/9308265}{{\ttfamily hep-ph/9308265}}].

\bibitem{Reuter:1993kw}
M.~Reuter and C.~Wetterich, \emph{{Effective average action for gauge theories
  and exact evolution equations}},
  \href{https://doi.org/10.1016/0550-3213(94)90543-6}{\emph{Nucl. Phys.}
  {\bfseries B417} (1994) 181}.

\bibitem{Ellwanger:1993mw}
U.~Ellwanger, \emph{{Proceedings, Workshop on Quantum field theoretical aspects
  of high energy physics: Bad Frankenhausen, Germany, September 20-24, 1993}},
  \href{https://doi.org/10.1007/BF01555911}{\emph{Z. Phys.} {\bfseries C62}
  (1994) 503} [\href{https://arxiv.org/abs/hep-ph/9308260}{{\ttfamily
  hep-ph/9308260}}].

\bibitem{Morris:1998da}
T.~R. Morris, \emph{{Nonperturbative QCD: Structure of the QCD vacuum:
  Proceedings, Yukawa International Seminar, YKIS'97, Kyoto, Japan, December
  2-12, 1997}}, \href{https://doi.org/10.1143/PTPS.131.395}{\emph{Prog. Theor.
  Phys. Suppl.} {\bfseries 131} (1998) 395}
  [\href{https://arxiv.org/abs/hep-th/9802039}{{\ttfamily hep-th/9802039}}].

\bibitem{Berges:2000ew}
J.~Berges, N.~Tetradis and C.~Wetterich, \emph{{Nonperturbative renormalization
  flow in quantum field theory and statistical physics}},
  \href{https://doi.org/10.1016/S0370-1573(01)00098-9}{\emph{Phys. Rept.}
  {\bfseries 363} (2002) 223}
  [\href{https://arxiv.org/abs/hep-ph/0005122}{{\ttfamily hep-ph/0005122}}].

\bibitem{Aoki:2000wm}
K.~Aoki, \emph{{Introduction to the nonperturbative renormalization group and
  its recent applications}},
  \href{https://doi.org/10.1142/S0217979200000923}{\emph{Int.J.Mod.Phys.}
  {\bfseries B14} (2000) 1249}.

\bibitem{Bagnuls:2000ae}
C.~Bagnuls and C.~Bervillier, \emph{{Exact renormalization group equations. An
  Introductory review}},
  \href{https://doi.org/10.1016/S0370-1573(00)00137-X}{\emph{Phys. Rept.}
  {\bfseries 348} (2001) 91}
  [\href{https://arxiv.org/abs/hep-th/0002034}{{\ttfamily hep-th/0002034}}].

\bibitem{Polonyi:2001se}
J.~Polonyi, \emph{{Lectures on the functional renormalization group method}},
  \href{https://doi.org/10.2478/BF02475552}{\emph{Central Eur. J. Phys.}
  {\bfseries 1} (2003) 1}
  [\href{https://arxiv.org/abs/hep-th/0110026}{{\ttfamily hep-th/0110026}}].

\bibitem{Pawlowski:2005xe}
J.~M. Pawlowski, \emph{{Aspects of the functional renormalisation group}},
  \href{https://doi.org/10.1016/j.aop.2007.01.007}{\emph{Annals Phys.}
  {\bfseries 322} (2007) 2831}
  [\href{https://arxiv.org/abs/hep-th/0512261}{{\ttfamily hep-th/0512261}}].

\bibitem{Gies:2006wv}
H.~Gies, \emph{{Introduction to the functional RG and applications to gauge
  theories}},
  \href{https://doi.org/10.1007/978-3-642-27320-9_6}{\emph{Lect.Notes Phys.}
  {\bfseries 852} (2012) 287}
  [\href{https://arxiv.org/abs/hep-ph/0611146}{{\ttfamily hep-ph/0611146}}].

\bibitem{Delamotte:2007pf}
B.~Delamotte, \emph{{An Introduction to the nonperturbative renormalization
  group}}, \href{https://doi.org/10.1007/978-3-642-27320-9_2}{\emph{Lect. Notes
  Phys.} {\bfseries 852} (2012) 49}
  [\href{https://arxiv.org/abs/cond-mat/0702365}{{\ttfamily
  cond-mat/0702365}}].

\bibitem{Sonoda:2007av}
H.~Sonoda, \emph{{The Exact Renormalization Group: Renormalization theory
  revisited}},  9, 2007, \href{https://arxiv.org/abs/0710.1662}{{\ttfamily
  0710.1662}}.

\bibitem{Igarashi:2009tj}
Y.~Igarashi, K.~Itoh and H.~Sonoda, \emph{{Realization of Symmetry in the ERG
  Approach to Quantum Field Theory}},
  \href{https://doi.org/10.1143/PTPS.181.1}{\emph{Prog. Theor. Phys. Suppl.}
  {\bfseries 181} (2010) 1} [\href{https://arxiv.org/abs/0909.0327}{{\ttfamily
  0909.0327}}].

\bibitem{Rosten:2010vm}
O.~J. Rosten, \emph{{Fundamentals of the Exact Renormalization Group}},
  \href{https://doi.org/10.1016/j.physrep.2011.12.003}{\emph{Phys. Rept.}
  {\bfseries 511} (2012) 177}
  [\href{https://arxiv.org/abs/1003.1366}{{\ttfamily 1003.1366}}].

\bibitem{Braun:2011pp}
J.~Braun, \emph{{Fermion Interactions and Universal Behavior in Strongly
  Interacting Theories}},
  \href{https://doi.org/10.1088/0954-3899/39/3/033001}{\emph{J. Phys.}
  {\bfseries G39} (2012) 033001}
  [\href{https://arxiv.org/abs/1108.4449}{{\ttfamily 1108.4449}}].

\bibitem{Nandori:2010ij}
I.~Nandori, \emph{{Bosonization and Functional Renormalization Group Approach
  in the Framework of QED$_{2}$}},
  \href{https://doi.org/10.1103/PhysRevD.84.065024}{\emph{Phys. Rev. D}
  {\bfseries 84} (2011) 065024}
  [\href{https://arxiv.org/abs/1008.2934}{{\ttfamily 1008.2934}}].

\bibitem{Daviet:2021whj}
R.~Daviet and N.~Dupuis, \emph{{Flowing bosonization in the nonperturbative
  functional renormalization-group approach}},
  \href{https://doi.org/10.21468/SciPostPhys.12.3.110}{\emph{SciPost Phys.}
  {\bfseries 12} (2022) 110}
  [\href{https://arxiv.org/abs/2111.11458}{{\ttfamily 2111.11458}}].

\bibitem{Zinn-Justin:1991ksq}
J.~Zinn-Justin, \emph{{Four fermion interaction near four-dimensions}},
  \href{https://doi.org/10.1016/0550-3213(91)90043-W}{\emph{Nucl. Phys. B}
  {\bfseries 367} (1991) 105}.

\bibitem{Karkkainen:1993ef}
L.~Karkkainen, R.~Lacaze, P.~Lacock and B.~Petersson, \emph{{Critical behavior
  of the three-dimensional Gross-Neveu and Higgs-Yukawa models}},
  \href{https://doi.org/10.1016/0550-3213(94)90309-3}{\emph{Nucl. Phys. B}
  {\bfseries 415} (1994) 781}
  [\href{https://arxiv.org/abs/hep-lat/9310020}{{\ttfamily hep-lat/9310020}}].

\bibitem{Wilson:1971dc}
K.~G. Wilson and M.~E. Fisher, \emph{{Critical exponents in 3.99 dimensions}},
  \href{https://doi.org/10.1103/PhysRevLett.28.240}{\emph{Phys. Rev. Lett.}
  {\bfseries 28} (1972) 240}.

\bibitem{Flore:2012wh}
R.~Flore, \emph{{Renormalization of the Nonlinear O(3) Model with Theta-Term}},
  \href{https://doi.org/10.1016/j.nuclphysb.2013.01.021}{\emph{Nucl. Phys. B}
  {\bfseries 870} (2013) 444}
  [\href{https://arxiv.org/abs/1208.5948}{{\ttfamily 1208.5948}}].

\bibitem{Efremov:2021fub}
A.~N. Efremov and A.~Ran\c{c}on, \emph{{Nonlinear sigma models on constant
  curvature target manifolds: A functional renormalization group approach}},
  \href{https://doi.org/10.1103/PhysRevD.104.105003}{\emph{Phys. Rev. D}
  {\bfseries 104} (2021) 105003}
  [\href{https://arxiv.org/abs/2109.09364}{{\ttfamily 2109.09364}}].

\bibitem{Fukushima:2022zor}
K.~Fukushima, T.~Shimazaki and Y.~Tanizaki, \emph{{Exploring the
  \ensuremath{\theta}-vacuum structure in the functional renormalization group
  approach}}, \href{https://doi.org/10.1007/JHEP04(2022)040}{\emph{JHEP}
  {\bfseries 04} (2022) 040}
  [\href{https://arxiv.org/abs/2202.00375}{{\ttfamily 2202.00375}}].

\bibitem{Affleck:1987ch}
I.~Affleck and F.~D.~M. Haldane, \emph{{Critical Theory of Quantum Spin
  Chains}}, \href{https://doi.org/10.1103/PhysRevB.36.5291}{\emph{Phys. Rev. B}
  {\bfseries 36} (1987) 5291}.

\bibitem{Altland:2006si}
A.~Altland and B.~Simons, \emph{{Condensed Matter Field Theory}}. Cambridge
  University Press, 8, 2023,
  \href{https://doi.org/10.1017/9781108781244}{10.1017/9781108781244}.

\bibitem{Mitter:1988xa}
P.~K. Mitter and T.~R. Ramadas, \emph{{The Two-Dimensional O(n) Nonlinear Sigma
  Model: Renormalization and Effective Actions}},
  \href{https://doi.org/10.1007/BF01256494}{\emph{Commun. Math. Phys.}
  {\bfseries 122} (1989) 575}.

\bibitem{Gawedzki:1986dt}
K.~Gawedzki and A.~Kupiainen, \emph{{Continuum Limit of the Hierarchical O($N$)
  Nonlinear $\sigma$ Model}},
  \href{https://doi.org/10.1007/BF01463394}{\emph{Commun. Math. Phys.}
  {\bfseries 106} (1986) 533}.

\bibitem{Flore:2012ma}
R.~Flore, A.~Wipf and O.~Zanusso, \emph{{Functional renormalization group of
  the non-linear sigma model and the $O(N)$ universality class}},
  \href{https://doi.org/10.1103/PhysRevD.87.065019}{\emph{Phys. Rev. D}
  {\bfseries 87} (2013) 065019}
  [\href{https://arxiv.org/abs/1207.4499}{{\ttfamily 1207.4499}}].

\bibitem{Percacci:2013jpa}
R.~Percacci and M.~Safari, \emph{{Functional renormalization of N scalars with
  O(N) invariance}},
  \href{https://doi.org/10.1103/PhysRevD.88.085007}{\emph{Phys. Rev. D}
  {\bfseries 88} (2013) 085007}
  [\href{https://arxiv.org/abs/1306.3918}{{\ttfamily 1306.3918}}].

\bibitem{Litim:2001up}
D.~F. Litim, \emph{{Optimized renormalization group flows}},
  \href{https://doi.org/10.1103/PhysRevD.64.105007}{\emph{Phys. Rev.}
  {\bfseries D64} (2001) 105007}
  [\href{https://arxiv.org/abs/hep-th/0103195}{{\ttfamily hep-th/0103195}}].

\bibitem{Reuter:1996cp}
M.~Reuter, \emph{{Nonperturbative evolution equation for quantum gravity}},
  \href{https://doi.org/10.1103/PhysRevD.57.971}{\emph{Phys. Rev. D} {\bfseries
  57} (1998) 971} [\href{https://arxiv.org/abs/hep-th/9605030}{{\ttfamily
  hep-th/9605030}}].

\bibitem{Souma:1999at}
W.~Souma, \emph{{Nontrivial ultraviolet fixed point in quantum gravity}},
  \href{https://doi.org/10.1143/PTP.102.181}{\emph{Prog. Theor. Phys.}
  {\bfseries 102} (1999) 181}
  [\href{https://arxiv.org/abs/hep-th/9907027}{{\ttfamily hep-th/9907027}}].

\bibitem{Oda:2015sma}
K.-y. Oda and M.~Yamada, \emph{{Non-minimal coupling in
  Higgs\textendash{}Yukawa model with asymptotically safe gravity}},
  \href{https://doi.org/10.1088/0264-9381/33/12/125011}{\emph{Class. Quant.
  Grav.} {\bfseries 33} (2016) 125011}
  [\href{https://arxiv.org/abs/1510.03734}{{\ttfamily 1510.03734}}].

\bibitem{Hamada:2017rvn}
Y.~Hamada and M.~Yamada, \emph{{Asymptotic safety of higher derivative quantum
  gravity non-minimally coupled with a matter system}},
  \href{https://doi.org/10.1007/JHEP08(2017)070}{\emph{JHEP} {\bfseries 08}
  (2017) 070} [\href{https://arxiv.org/abs/1703.09033}{{\ttfamily
  1703.09033}}].

\bibitem{DiFrancesco:1997nk}
P.~Di~Francesco, P.~Mathieu and D.~Senechal, \emph{{Conformal Field Theory}},
  Graduate Texts in Contemporary Physics. Springer-Verlag, New York, 1997,
  \href{https://doi.org/10.1007/978-1-4612-2256-9}{10.1007/978-1-4612-2256-9}.

\bibitem{Codello:2013iqa}
A.~Codello, G.~D'Odorico and C.~Pagani, \emph{{A functional RG equation for the
  c-function}}, \href{https://doi.org/10.1007/JHEP07(2014)040}{\emph{JHEP}
  {\bfseries 07} (2014) 040} [\href{https://arxiv.org/abs/1312.7097}{{\ttfamily
  1312.7097}}].

\bibitem{Sonoda:2020vut}
H.~Sonoda and H.~Suzuki, \emph{{Gradient flow exact renormalization group}},
  \href{https://doi.org/10.1093/ptep/ptab006}{\emph{PTEP} {\bfseries 2021}
  (2021) 023B05} [\href{https://arxiv.org/abs/2012.03568}{{\ttfamily
  2012.03568}}].

\bibitem{Makino:2018rys}
H.~Makino, O.~Morikawa and H.~Suzuki, \emph{{Gradient flow and the Wilsonian
  renormalization group flow}},
  \href{https://doi.org/10.1093/ptep/pty050}{\emph{PTEP} {\bfseries 2018}
  (2018) 053B02} [\href{https://arxiv.org/abs/1802.07897}{{\ttfamily
  1802.07897}}].

\bibitem{Sonoda:2019ibh}
H.~Sonoda and H.~Suzuki, \emph{{Derivation of a gradient flow from the exact
  renormalization group}},
  \href{https://doi.org/10.1093/ptep/ptz020}{\emph{PTEP} {\bfseries 2019}
  (2019) 033B05} [\href{https://arxiv.org/abs/1901.05169}{{\ttfamily
  1901.05169}}].

\bibitem{Miyakawa:2021hcx}
Y.~Miyakawa and H.~Suzuki, \emph{{Gradient flow exact renormalization group:
  Inclusion of fermion fields}},
  \href{https://doi.org/10.1093/ptep/ptab100}{\emph{PTEP} {\bfseries 2021}
  (2021) 083B04} [\href{https://arxiv.org/abs/2106.11142}{{\ttfamily
  2106.11142}}].

\bibitem{Miyakawa:2021wus}
Y.~Miyakawa, H.~Sonoda and H.~Suzuki, \emph{{Manifestly gauge invariant exact
  renormalization group for quantum electrodynamics}},
  \href{https://doi.org/10.1093/ptep/ptac003}{\emph{PTEP} {\bfseries 2022}
  (2022) 023B02} [\href{https://arxiv.org/abs/2111.15529}{{\ttfamily
  2111.15529}}].

\bibitem{Sonoda:2022fmk}
H.~Sonoda and H.~Suzuki, \emph{{One-particle irreducible Wilson action in the
  gradient flow exact renormalization group formalism}},
  \href{https://doi.org/10.1093/ptep/ptac047}{\emph{PTEP} {\bfseries 2022}
  (2022) 053B01} [\href{https://arxiv.org/abs/2201.04448}{{\ttfamily
  2201.04448}}].

\bibitem{Miyakawa:2023yob}
Y.~Miyakawa, H.~Sonoda and H.~Suzuki, \emph{{Chiral anomaly as a composite
  operator in the gradient flow exact renormalization group formalism}},
  \href{https://doi.org/10.1093/ptep/ptad074}{\emph{PTEP} {\bfseries 2023}
  (2023) 063B03} [\href{https://arxiv.org/abs/2304.14753}{{\ttfamily
  2304.14753}}].

\bibitem{Abe:2022smm}
Y.~Abe, Y.~Hamada and J.~Haruna, \emph{{Fixed point structure of the gradient
  flow exact renormalization group for scalar field theories}},
  \href{https://doi.org/10.1093/ptep/ptac021}{\emph{PTEP} {\bfseries 2022}
  (2022) 033B03} [\href{https://arxiv.org/abs/2201.04111}{{\ttfamily
  2201.04111}}].

\bibitem{Haruna:2023spq}
J.~Haruna and M.~Yamada, \emph{{Gradient Flow Exact Renormalization Group for
  Scalar Quantum Electrodynamics}},
  \href{https://arxiv.org/abs/2312.15673}{{\ttfamily 2312.15673}}.

\bibitem{Makino:2014sta}
H.~Makino and H.~Suzuki, \emph{{Renormalizability of the gradient flow in the
  2D $O(N)$ non-linear sigma model}},
  \href{https://doi.org/10.1093/ptep/ptv028}{\emph{PTEP} {\bfseries 2015}
  (2015) 033B08} [\href{https://arxiv.org/abs/1410.7538}{{\ttfamily
  1410.7538}}].

\bibitem{Makino:2014cxa}
H.~Makino, F.~Sugino and H.~Suzuki, \emph{{Large-$N$ limit of the gradient flow
  in the 2D $O(N)$ nonlinear sigma model}},
  \href{https://doi.org/10.1093/ptep/ptv044}{\emph{PTEP} {\bfseries 2015}
  (2015) 043B07} [\href{https://arxiv.org/abs/1412.8218}{{\ttfamily
  1412.8218}}].

\bibitem{Aoki:2014dxa}
S.~Aoki, K.~Kikuchi and T.~Onogi, \emph{{Gradient Flow of O(N) nonlinear sigma
  model at large N}},
  \href{https://doi.org/10.1007/JHEP04(2015)156}{\emph{JHEP} {\bfseries 04}
  (2015) 156} [\href{https://arxiv.org/abs/1412.8249}{{\ttfamily 1412.8249}}].

\bibitem{Gies:2001nw}
H.~Gies and C.~Wetterich, \emph{{Renormalization flow of bound states}},
  \href{https://doi.org/10.1103/PhysRevD.65.065001}{\emph{Phys. Rev. D}
  {\bfseries 65} (2002) 065001}
  [\href{https://arxiv.org/abs/hep-th/0107221}{{\ttfamily hep-th/0107221}}].

\bibitem{Fukushima:2021ctq}
K.~Fukushima, J.~M. Pawlowski and N.~Strodthoff, \emph{{Emergent hadrons and
  diquarks}}, \href{https://doi.org/10.1016/j.aop.2022.169106}{\emph{Annals
  Phys.} {\bfseries 446} (2022) 169106}
  [\href{https://arxiv.org/abs/2103.01129}{{\ttfamily 2103.01129}}].

\bibitem{Fukushima:2023wnl}
K.~Fukushima, J.~Horak, J.~M. Pawlowski, N.~Wink and C.~P. Zelle, \emph{{The
  nuclear liquid-gas transition in QCD}},
  \href{https://arxiv.org/abs/2308.16594}{{\ttfamily 2308.16594}}.

\end{thebibliography}\endgroup
\end{document}